\documentclass[aps,preprint]{revtex4}
\usepackage[dvips]{graphics,graphicx}
\usepackage{amsmath}

\begin{document}
\title{Induced tunneling and localization for a quantum particle in tilted two-dimensional lattices}
\author{Evgeny~N.~Bulgakov$^{1}$ and Andrey~R.~Kolovsky$^{1,2}$}
\affiliation{$^1$Kirensky Institute of Physics, 660036 Krasnoyarsk, Russia}
\affiliation{$^2$Siberian Federal University, 660041 Krasnoyarsk, Russia}
%{\small \em andrey.r.kolovsky@gmail.com}
\date{\today}

\begin{abstract}
We consider a quantum particle in tilted two-dimensional lattices in the tight-binding approximations. We found that for some lattice geometries and certain orientations of the static force with respect to the lattice primary axes the particle can freely move across the lattice in the direction perpendicular to the vector of the static force. This effect is argued to be analogue of the photon-induced tunneling in driven one-dimensional lattices. The obtained dispersion relation for the transverse  motion of the particle draws this analogy further by eventually showing band collapses when a control parameter is varied. 
\end{abstract}

%\pacs{03.65.-w}{Quantum mechanics}
%\pacs{03.75.Lm}{Tunneling, Josephson effect, Bose-Einstein condensates in periodic potentials, solitons, vortices, and topological excitations}
\maketitle

%%%%%%%%%%%%%%%%%%%%%%%%%%%%%%%%%
{\em 1.} Band collapse or dynamic localization is a destructive interference effect that occurs in periodically driven one-dimensional (1D) lattices. This effect has been of considerable theoretical interest since the seminal work by Dunlap and Kenkre \cite{Dunl86} and was experimentally observed for cold atoms in optical lattices \cite{Madi98,Lign07} and the light in coupled waveguide arrays \cite{Long06}. It takes place for certain amplitudes of the driving force $F_\omega$ that are determined by roots of the equation ${\cal J}_0(aF_\omega/\hbar\omega)=0$, where ${\cal J}_0(z)$ is the zero-order Bessel function, $\omega$  the driving frequency, and $a$ the lattice period. One meets the same effect in tilted 1D lattices provided the Bloch frequency $\omega_B=aF/\hbar$ is a multiple of the driving frequency, i.e., $\omega_B=q\omega$ \cite{Sias08,Ivan08,Haller}. In this case the original dispersion relation $E(\kappa)=2t\cos(\kappa a)$ for the quantum particle in the 1D lattice is substituted by 
%********************************************************
\begin{equation}
\label{1}
E(\kappa)=t {\cal J}_q\left(\frac{aF_\omega}{\hbar\omega_B}\right) \cos(\kappa a) 
\end{equation} 
(here $t$ is the hopping matrix element and $\kappa$ the particle quasimomentum). Then the localization condition is given by roots of the $q$th order Bessel function \cite{Zhao91}. Notice that because ${\cal J}_q(z)\sim z^q$ for $z\rightarrow0$  the particle can be delocalized in the tilted lattice only in the presence of driving -- the phenomenon known as photon-induced or photon-assisted tunneling. Thus the resonant driving first induces the tunneling and then suppress it for higher amplitude of the driving force. 

In the present work we discuss emergence and suppression of tunneling for a quantum particle in tilted two-dimensional (2D) lattices \cite{remark0}.  Similar to tilted 1D lattices,  the necessary condition to observe these effects is the resonance condition, now on two Bloch frequencies associated with two components of the vector ${\bf F}$ of a static force.  This condition, however, is not sufficient and the effects are absent in simple lattices like the square or triangle lattices. The other requirement is that the 2D lattice should have nontrivial geometry and consist at least of two sub-lattices. In what follows we specifically address the geometry realized in the recent experiment \cite{Tarr12,Uehi13} -- a square lattice with three different hopping matrix elements, see Fig.~1. For this lattice the resonance condition reads $F_x/F_y=r/q$, where $r$ and $q$ are co-prime numbers and we refer to the coordinate system determined by the lattice primary axes. Our aim is to obtain an analogue of Eq.~(\ref{1}) where control parameters are magnitude of the static force $F=|{\bf F}|$ and two prime numbers $r$ and $q$. 
%%%%%%%%%%%%%%%%%%%%%%%%%%%%%%%%%%%%%%%%%%%%%%%%%%%
\begin{figure}[ht]
\center
\includegraphics[width=0.6\textwidth]{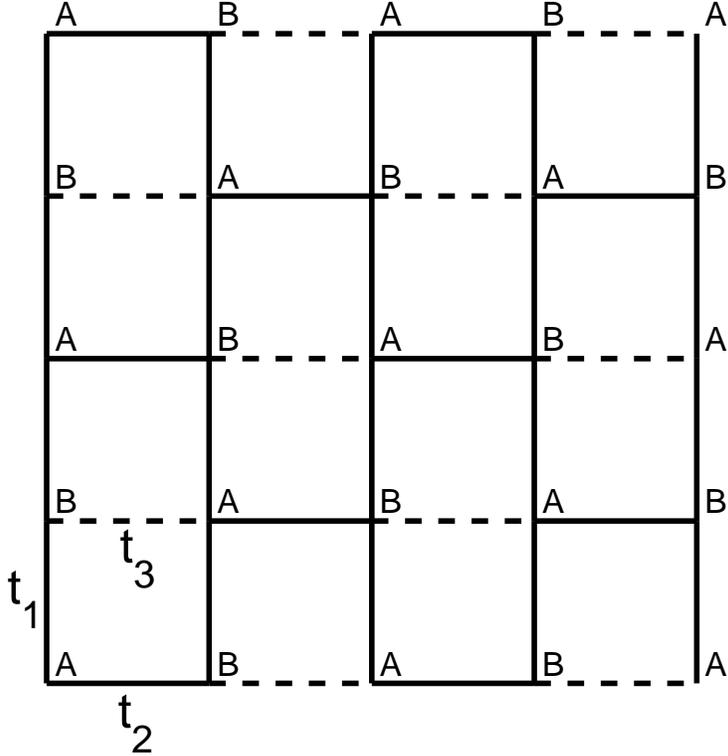}
\caption{The tight-binding model with three different hopping amplitudes.  For $t_3=t_2$ the model reduces to the simple square lattice while for $t_3=0$ it is topologically equivalent to the honeycomb lattice. Notice that primary axes of the square and honeycomb-like lattices are rotated by $\pi/4$ relative to each other.}
\label{fig1}
\end{figure}

%%%%%%%%%%%%%%%%%%%%%%%%%%%%%%%%%%%%%%%%%%%%%%
{\em 2.} Initial step of the analysis closely follows our recent work \cite{92} devoted to the Wannier-Stark states in the honeycomb lattice. First we introduce the new coordinate system $(\eta,\xi)$ which is rotated by the angle $\theta=\arctan(F_x/F_y)=\arctan(r/q)$ with respect to the lattice primary axes. In this coordinate system the Stark energy ${\bf F}{\bf r}$ depends only on $\xi$ and, hence, we can use the ansatz  $\Psi(\eta,\nu)=\exp(i\kappa\eta)\psi(\xi)$ to separate the variables \cite{Naka93}. Notice that in new variables the coordinates ${\bf r}_i$ of the lattice sites are multiple of the period  
%********************************************************
\begin{equation}
\label{0}
d=\frac{\sqrt{2}a}{\sqrt{r^2+q^2}} \;, 
\end{equation} 
where $a$ is the period of the square lattice with no sub-lattices ($t_3=t_2$). Using the above ansatz the stationary Schr\"odingier equation reduces to the following system of coupled algebraic equations 
%****************************************************
\begin{equation}
\label{2}
\begin{split}
&E\psi^{A}_{j}=-t_1e^{-ir\kappa d}\psi^{B}_{j-q}-t_1e^{iq\kappa d}\psi^{B}_{j-r}
\\
&-t_2e^{i(q-r)\kappa d}\psi^{B}_{j-q-r}-t_3\psi^{B}_{j} + Fdj \psi^{A}_{j} \;,
\\
&E\psi^{B}_{j}=-t_1e^{ir\kappa d}\psi^{A}_{j+q}-t_1e^{-iq\kappa d}\psi^{A}_{j+r}
\\
&-t_2e^{i(r-q)\kappa d}\psi^{A}_{j+r+q}-t_3\psi^{A}_{j} + (Fdj+E_0)\psi^{B}_{j} \;,
\end{split}
\end{equation}
where $E_0=Fd(r+q)/2$ and $A$ and $B$ are the sub-lattice indexes.  As an example Fig.~2 shows numerical solution of (\ref{2}) for $(r,q)=(2,1)$, $(t_1,t_2,t_3)=(1,0.5,0.25)$, and $F=2.3$. It is seen in Fig.~2 that the energy bands form two Wannier-Stark ladders which are related to each other by a reflection symmetry. Namely, given $E(\kappa)$ the dispersion relation for energy bands of the first ladder, the dispersion relation for energy bands of the second ladder is $-E(\kappa)$ if $(r+q)/2$ is an integer and $-E(\kappa)+Fd/2$ if  $(r+q)/2$ is a half-integer.

To approach Eq.~(\ref{2}) analytically we Fourier transform it. This substitutes algebraic equations by ordinary differential equations. Denoting by $\tilde{\psi}^{A,B}(\chi)$ the Fourier images of the functions $\psi^{A,B}(\xi)=\sum_j  \psi^{A,B}_j \delta(\xi-dj)$ we have
%*******************************************************
\begin{equation}
\label{3}
\begin{split}
&
iF\frac{{\rm d}\tilde{\psi}^{A}}{{\rm d}\chi}=E\tilde{\psi}^{A} + \tilde{G}\widetilde{\psi}^{B} \;,
\\
&iF\frac{{\rm d}\tilde{\psi}^{B}}{{\rm d}\chi}=(E-E_0)\tilde{\psi}^{B}+\widetilde{G}^{*}\tilde{\psi}^{A} \;,
\end{split}
\end{equation}
where the coefficient $\widetilde{G}$ is a function of $\chi$ and $\kappa$, $\widetilde{G}(\chi,\kappa)=t_1\exp[-id(\kappa r+\chi q)]+t_1\exp[-id(\chi r-\kappa q)]+t_2\exp[-id\chi(r+q)+id\kappa(q-r)]+t_3$. It is convenient to eliminate the diagonal part in the right hand side of Eq.~(\ref{3}) and rewrite the function $\tilde{G}(\chi,\kappa)$ in the rotated variables $\alpha=\chi \cos\theta -\kappa \sin\theta$ and $\beta=\chi \sin\theta + \kappa \cos\theta$. We have
%****************************************************
\begin{equation}
\label{4}
i\frac{{\rm d}c_1 }{{\rm d}\chi}=\frac{1}{F} G c_2  \;,\quad
i\frac{{\rm d}c_2 }{{\rm d}\chi}=\frac{1}{F} G^{*}c_1  \;,
\end{equation}
where  $c_1=\tilde{\psi}^{A}\exp(iE\chi/F)$, $c_2=\tilde{\psi}^{B}\exp(iE\chi/F)\exp[-i(\alpha+\beta)/2]$, and 
%****************************************************
\begin{equation}
\label{5}
G=2t_1\cos\left(\frac{\alpha-\beta}{2}\right)+2t_2\cos\left(\frac{\alpha+\beta}{2}\right)+(t_3-t_2)\exp\left(i\frac{\alpha+\beta}{2}\right) \;.
\end{equation}
Since the coefficient (\ref{5}) is a periodic function of the variable $\chi$, we can use the Bogoliubov-Metropolskii technique \cite{book} to analyze Eq.~(\ref{4}). This method involves averaging the function $G$ and gives the solution of Eq.~(\ref{4}) in the form of a series over the parameter $\epsilon=1/F$. 

Before proceeding with the above mentioned method we show that for $t_3=t_2$ Eq.~(\ref{4}) has trivial solution that corresponds to flat bands of the simple square lattice \cite{remark1},
%****************************************************
\begin{equation}
\label{6}
E_p(\kappa)=F\frac{a}{\sqrt{r'^2+q'^2}} p \;,\quad  \frac{F'_x}{F'_y}=\frac{r'}{q'}  \;,\quad  p=0,\pm 1,\ldots
\end{equation}
(here $F'_x$ and $F'_y$ are components of ${\bf F}$ in the coordinate system determined by the primary axes of the square lattice).    Let us denote by $G_R$ and $G_I$ the real and imaginary parts of the function (\ref{5}). After the substitution $u=(c_1 +c_2 ) \exp(i\epsilon\int G_R {\rm d}\chi)$ and $v=(c_1 -c_2 ) \exp(-i\epsilon\int G_R {\rm d}\chi)$ Eq.~(\ref{4}) takes the form
%****************************************************
\begin{equation}
\label{7}
i\frac{{\rm d}u}{{\rm d}\chi}=\epsilon X v \;,\quad
i\frac{{\rm d}v}{{\rm d}\chi}=\epsilon X^{*} u \;,
\end{equation}
where $X=-iG_I\exp(2i\epsilon \int G_R {\rm d}\chi)$. Since for $t_3=t_2$ the imaginary part of the function (\ref{5}) vanishes, the solution of (\ref{7}) is the constant function $(u,v)^T=(u_0,v_0)^T$. Then the energy spectrum (\ref{6}) follows from the requirement (quantization condition) that the functions $\tilde{\psi}^{A,B}=\tilde{\psi}^{A,B}(\chi;\kappa)$ are periodic function of $\chi$ with the period $2\pi/d$.

We proceed with general case $t_3\ne t_2$. Restricting ourselves by the second order over the parameter $\epsilon=1/F$, the Bogoliubov equation for the column function $(u,v)^T$ reads
%*******************************************************
\begin{equation}
\label{8} 
i\frac{{\rm d}}{{\rm d} \chi}\left(
\begin{array}{c}
  u\\ v
\end{array}\right)
=\epsilon\begin{pmatrix}
  0 & \langle X\rangle \\
  \langle X^*\rangle & 0
\end{pmatrix} \begin{pmatrix}
  u \\ v
\end{pmatrix}
+\epsilon^2\begin{pmatrix}
  \langle-iX'^{*}X\rangle & 0 \\
  0 &  \langle-iX'X^{*}\rangle
\end{pmatrix}\begin{pmatrix}
  u\\ v
\end{pmatrix}  \;,
\end{equation}
where angular brackets denote the average over the period $2\pi/d$ and the prime sign is a shortcut for integral from oscillating part of $X$, i.e., $X'(\chi)=\sum_{\nu\ne0}\frac{\exp(i\nu\chi)}{i\nu} X_\nu$.  The solution of (\ref{8}) is $(u,v)^T=\exp(-i\epsilon\lambda \chi) (u_0,v_0)^T$ where 
%****************************************************
\begin{equation}
\label{9}
\lambda=\pm \sqrt{|\langle X\rangle|^2 + \epsilon^2\langle-iX'X^*\rangle^2 }
\end{equation}
(notice that $\langle-iX'X^*\rangle$ is real). The meaning of the quantity $\lambda=\lambda(\kappa)$ is the correction to the energy spectrum (\ref{6}). Using explicit form of $X$ this correction can be expressed in terms of the Bessel functions.  For the mean value of $X$ we have
%****************************************************
\begin{equation}
\label{10}
 \langle X\rangle=
  \begin{cases}
    -i(t_3-t_2) {\cal J}_{m}(z_1) {\cal J}_{n}(z_2)\sin\left[\kappa d\frac{r^2+q^2}{r-q}(1+n)\right] \;,& (n+m) \text{ is even}, \\
    -(t_3-t_2) {\cal J}_{m}(z_1) {\cal J}_{n}(z_2)\cos\left[\kappa d\frac{r^2+q^2}{r-q}(1+n)\right] \;,  & (n+m) \text{ is odd} \;,
  \end{cases}
 \end{equation}
where integer numbers $n$ and $m$ satisfy the equation $(r-q)m=-(r+q)(1+n)$ and arguments of the Bessel functions are 
%***************************************************
\begin{equation}
\label{11} 
z_1=\frac{8t_1}{Fd(r-q)} \;,\quad  z_2=\frac{4(t_2+t_3)}{Fd(r+q)} \;.
\end{equation}
Analogously, the mean $\langle-iX'X^*\rangle$ is given by product of four Bessel functions \cite{Sup}. Equations (\ref{9}-\ref{11}) constitute the main result of the paper. As shown below, these equations provide an estimate for the maximal width of the bands and describe asymptotic behavior of the energy bands in the limit $F\rightarrow\infty$.
%#########################################
\begin{figure}[b]
\center
\includegraphics[width=0.8\textwidth]{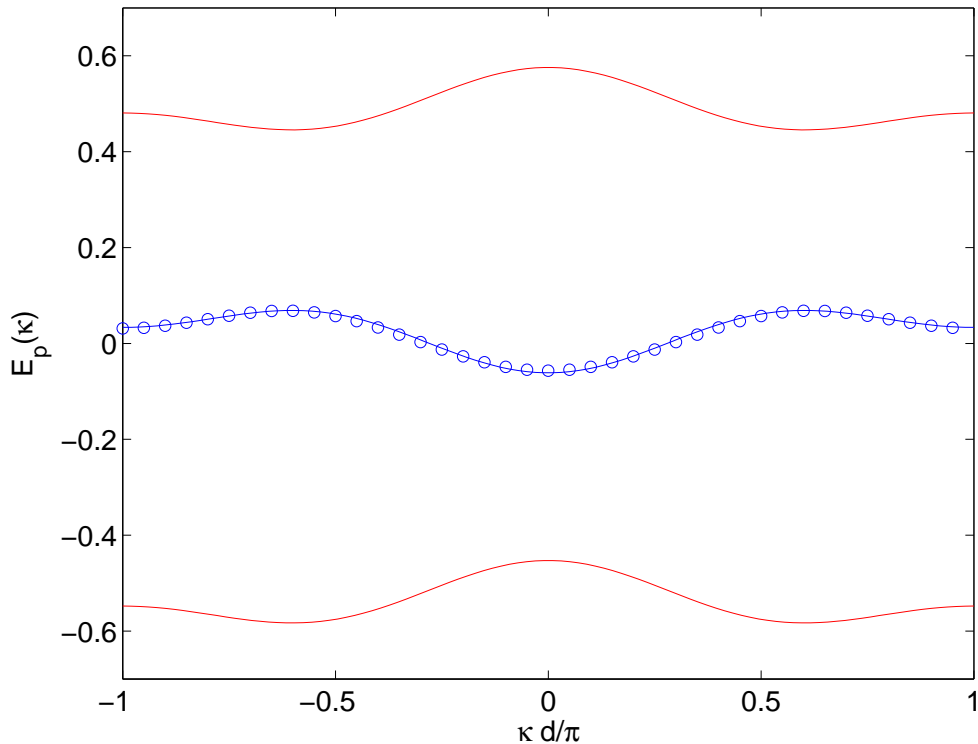}
\caption{Energy spectrum for $F=2.3$. The other parameters are $F_x/F_y=r/q=2/1$, and $(t_1,t_2,t_3)=(1,0.5,0.25)$. Open circles are the analytical result according to Eq.~(\ref{12}).}
\label{fig2}
\end{figure}

%%%%%%%%%%%%%%%%%%%%%%%%%%%%%%%%%%%%%%%%%%%%%%
{\em 3.} Let us consider as a generic example the case $(r,q)=(2,1)$. For this direction of the static force Eq.~(\ref{10}) takes the form
%*************************************************
\begin{equation}
\label{12}
\begin{split}
 & \langle X\rangle=(t_3-t_2)\left[{\cal J}_0(z_1){\cal J}_1(z_2)+{\cal J}_3(z_1){\cal J}_0(z_2)\cos(5\kappa d) \right. \\
 &\left. -{\cal J}_3(z_1){\cal J}_2(z_2)\cos(5\kappa d)- {\cal J}_6(z_1){\cal J}_1(z_2)\cos(10\kappa d)  + \ldots \right] \;.
\end{split}
\end{equation}
Notice that arguments of the Bessel functions are proportional to $1/F$. Thus different terms in the square brackets have different asymptotic if $F\rightarrow\infty$. In Eq.~(\ref{12}) we keep all terms up to the 7th power of $1/F$ and we checked that the next order Bogoliubov correction does not contain terms larger than $(1/F)^8$. Open circles in Fig.~2 show the dispersion relation calculated by using Eq.~(\ref{12}). Nice correspondence with the exact numerical result is noticed. Figure 3 shows the width of the depicted energy bands as the function of $F$. It follows from (\ref{9}) and (\ref{12}) that for $F\rightarrow\infty$ the width decreases as $1/F^3$. In the opposite limit the width takes its maximal value at $F\approx4.5$. Remarkably, already the first term with $\kappa$-dependence in the series (\ref{12}) provides accurate estimate for this maximal value (compare the solid and dashed lines in Fig.~3).
%#######################################
\begin{figure}[t]
\center
\includegraphics[width=0.8\textwidth]{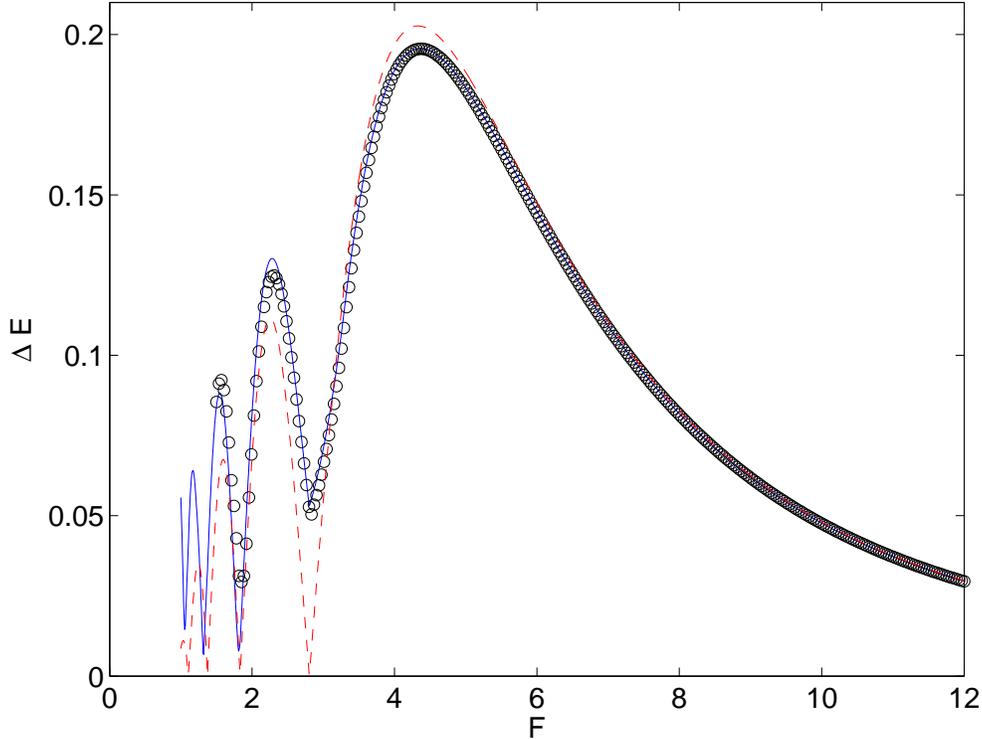}
\caption{Width of the energy bands as the function of $F$ for parameters of Fig.~2. The dashed and solid lines are plotted by keeping in Eq.~(\ref{12}) the first two and four terms, respectively. Symbols are the exact numerical result.}
\label{fig3}
\end{figure}

\begin{figure}
\center
\includegraphics[width=0.8\textwidth]{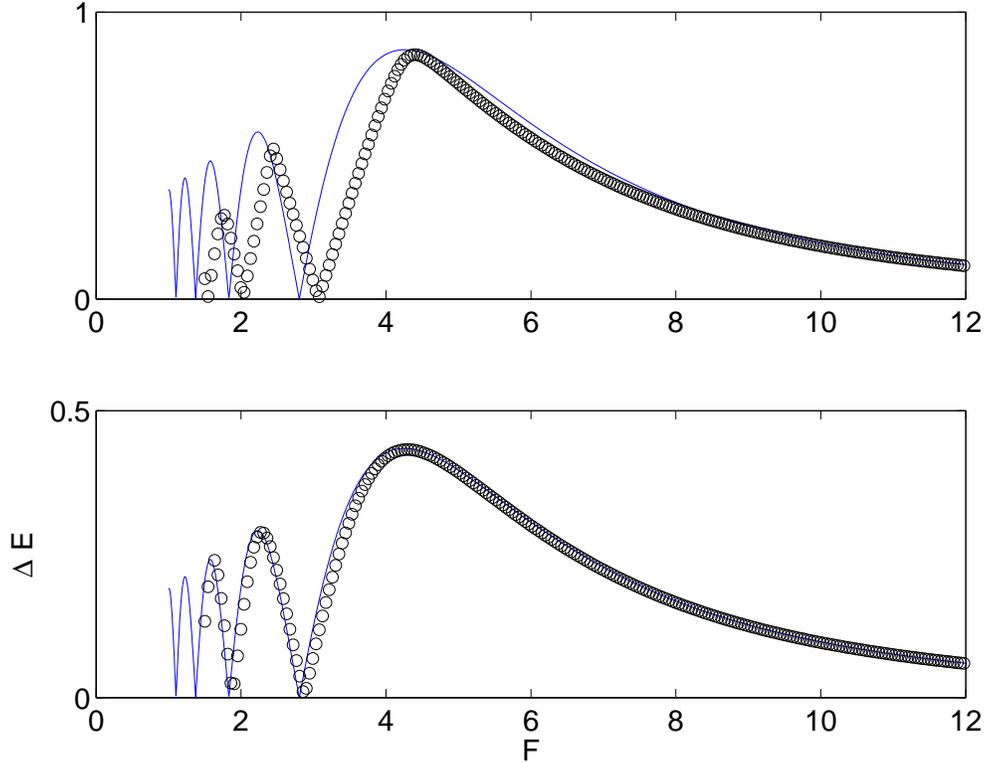}
\caption{Width of the energy bands as the function of $F$ for $(r,q)=(2,1)$ and $(t_1,t_2,t_3)=(1,0.25,-0.25)$, lower panel, and $(t_1,t_2,t_3)=(1,0.5,-0.5)$, upper panel.}
\label{fig4}
\end{figure}

Next we discuss an interesting case $t_3=-t_2$, which can be viewed as a charged particle in a square lattice in the presence of a staggered magnetic field with $\pi$-flux through the elementary cell. The system of this kind can be realized with cold atoms in a square optical lattice by properly driving the atoms by additional laser beams \cite{Adel11}. As follows from (\ref{11}), for $t_3=-t_2$ the argument $z_2$ equals to zero, that essentially simplifies all equations. For example, in the above considered case $(r,q)=(2,1)$ the first Bogoliubov approximation contains only one term and the dispersion relation reads
%***************************************************
\begin{equation}
\label{13} 
E(\kappa)=-2t_3 {\cal J}_3\left(\frac{8t_1}{F d}\right) \cos(5\kappa d) \;.
\end{equation}
%
%Eq.~(\ref{13}) predicts band collapse for $F$ determined by zeros of the 3rd order Bessel function and numerical analysis of the energy spectrum fairly confirms this prediction \cite{Sup}.
Fig.~\ref{fig4} compares Eq.~(\ref{13}) with numerical results for two sets of hopping amplitudes: $(t_1,t_2,t_3)=(1,0.25,-0.25)$ and $(t_1,t_2,t_3)=(1,0.5,-0.5)$. Almost complete band collapses are clearly seen in the figure. Comparing the upper and lower panels we also conclude that the actual parameter of the perturbation theory is $|t_2-t_3|/F\equiv \gamma/F$ but not just $1/F$. In general, the smaller $\gamma$ the further we can go in the region of small $F$.
%********************************************************

%%%%%%%%%%%%%%%%%%%%%%%%%%%%%%%%%%%%%%%%%%%%%%
{\em 4.} In conclusion, we analyzed quantum particle in tilted 2D lattices with square symmetry for orientations of the static force ${\bf F}$ given by the rationality condition $F_x/F_y=r/q$. It is shown that for these orientations the system has common features with the other fundamental problem -- the particle in periodically driven 1D lattices. Namely, the energy bands of both systems show non-monotonic behavior when a control parameter (for example, the static force $F$) is varied,  with partial or complete band collapse. We developed analytical method which provides explicit expression for the particle dispersion relation. The reported results can be verified in present-day laboratory experiments with cold atoms in 2D optical lattices by studying the ballistic spreading of atoms. Finally, we remark that the presented analysis can be also viewed as theory of the 2D Wannier-Stark states that remain an intriguing problem of the single-particle quantum mechanics.

The authors express their gratitude to D.~N.~Maksimov for useful remarks and acknowledge financial support of Russian Academy of Sciences through the SB RAS integration project  No.29 {\em Dynamics of atomic Bose-Einstein condensates in optical lattices} and the RFBR project No.12-02-00094 {\em Tunneling of the macroscopic quantum states}.

%%%%%%%%%%%%%%%%%%%%%%%%%%%%%%%%%%%%%%%%%%%%%%%%%%%%

%%%%%%%%%%%%%%%%%%%%%%%%%%%%%%%%
\newpage
\section{Appendix}

\subsection{The case $r=q$}

If $t_3=t_2$ and the static force is aligned with vertical or horizontal bonds (i.e., $r'$ or $q'$ equals to zero) Eq.~(\ref{6}) takes the form 
%*********************************************************
\begin{equation}
\label{a1}
E_p(\kappa)=aFp-2t_{1,2}\cos(\kappa a) \;.
\end{equation}
Notice that the energy bands have a finite width which is independent of $F$. If $t_3\ne t_2$  we meet a similar situation for $(r,q)=(1,1)$ where the particle tunnels along the vertical bonds. Including the first-order correction the dispersion relation reads
%*********************************************************
\begin{equation}
\label{a2}
E(\kappa)=\pm\sqrt{(t_2-t_3)^2{\cal J}_1^2(z)+4t_1^2\cos^2(\kappa d)} \;,\quad z=2(t_1+t_3)/Fd \;,
\end{equation}
and is seen to coincide with (\ref{a1}) in the limit of large $F$. 

The case $(r,q)=(1,-1)$ is more involved. In this case Eq.~(\ref{a2}) is substituted by 
%*********************************************************
\begin{equation}
\label{a3}
E(\kappa)=\pm\sqrt{(t_2-t_3)^2{\cal J}_0^2(z)\sin^2(\kappa d)+(t_2+t_2)^2\cos^2(\kappa d)} \;,\quad z=4t_1/Fd \;.
\end{equation}
It follows from (\ref{a3}) that $E(\kappa)$ may have different asymptotic depending on the hopping amplitude $t_3$. Namely, if $t_3=t_2$ (simple square lattice) we recover the dispersion relation (\ref{a1}), while for $t_3=0$ (honeycomb-like lattice) we have 
%*********************************************************
\begin{equation}
\label{a4}
E_p(\kappa)\approx t_2\left(1-\frac{4t_1^2}{F^2}{d^2}\sin^2(\kappa d)\right) 
\end{equation}
where the band width decreases as $1/F^2$. 

%Let us also mention an interesting case $t_3=-t_2$ where the band width increases when $F$ is increased,
%%*********************************************************
%\begin{equation}
%\label{a5}
%E(\kappa)=\pm t_2{\cal J}_0\left( \frac{4t_1}{Fd} \right)\sin(\kappa d) \;.
%\end{equation}
%%
%%###########################################
%\begin{figure}
%\center
%\includegraphics[width=0.8\textwidth]{fig4.eps}
%\caption{Width of the energy bands as the function of $F$ for $(t_1,t_2,t_3)=(1,0.5,0.25)$ and $(r,q)=(1,1)$, upper panel, and $(r,q)=(1,-1)$, lower panel.}
%\label{fig4}
%\end{figure}

%%%%%%%%%%%%%%%%%%%%%%%%%%
\subsection{Second order corrections}
We present explicit form of the second-order correction given by the term $\langle -iX'X^*\rangle$. As it was mentioned in the main text, this correction is given by product of four Bessel functions with the indexes $n$, $m$, $n'$, and $m'$, respectively. We have two contributions. The first contribution is given by
%*******************************************************
\begin{equation}
\label{b1}
\begin{split}
& A=\sum_{\nu_{+}(n,m)=\nu_{-}(n',m')\neq 0}
 -\frac{(t_3-t_2)^2}{d\nu_{+}(n,m)}{\cal J}_{m}(z_1){\cal J}_{n}(z_2){\cal J}_{m'}(z_1){\cal J}_{n'}(z_2)\\
 &\cos[(\mu_{+}(n,m)-\mu_{-}(n',m'))\kappa d/2] \;,
 \end{split}
\end{equation}
where 
%********************************************************* 
\begin{equation}
\label{b2}
\begin{split}
  & \nu_{\pm}(m,n)=m(r-q)+(n\pm 1)(r+q),\\
  & \mu_{\pm}(m,n)=(n \pm 1)(r-q)-m(r+q) \;.
 \end{split}
\end{equation}
The second contribution refers only to the case where $n+n'+m+m'$ is an odd number and reads
%*********************************************************
\begin{equation}
\label{b3}
\begin{split}
 & A=\sum_{\nu_{+}(n,m)=\nu_{+}(n',m')\neq 0}
 \frac{(t_3-t_2)^2}{d\nu_{+}(n,m)}{\cal J}_{m}(z_1) {\cal J}_{n}(z_2){\cal J}_{m'}(z_1) {\cal J}_{n'}(z_2)\\
 &\cos[(\mu_{+}(n,m)-\mu_{+}(n',m'))\kappa d/2] \;.
\end{split}
\end{equation}
%

%%%%%%%%%%%%%%%%%%%%%%%%%%%%
%\subsection{The case $t_3=-t_2$}
%As mentioned in the main text, for $t_3=-t_2$ the first order correction is given by Eq.~(\ref{13}). Fig.~5 compares this equation with exact numerical results (symbols) for two sets of the hopping amplitudes: $(t_1,t_2,t_3)=(1,0.25,-0.25)$ and $(t_1,t_2,t_3)=(1,0.5,-0.5)$. Almost complete band collapses are clearly seen in the figure. Comparing the upper and lower panels we also conclude that actual parameter of the perturbation theory is $|t_2-t_3|/F\equiv \gamma/F$ but not just $1/F$. In general, the smaller $\gamma$ the further we can go in the region of small $F$.
%%********************************************************
%\begin{figure}
%\center
%\includegraphics[width=0.8\textwidth]{fig4.eps}
%\caption{Width of the energy bands as the function of $F$ for $(r,q)=(2,1)$ and $(t_1,t_2,t_3)=(1,0.25,-0.25)$, lower panel, and $(t_1,t_2,t_3)=(1,0.5,-0.5)$, upper panel. Symbols show the exact numerical result.}
%\label{fig5}
%\end{figure}


\begin{thebibliography}{10}

\bibitem{Dunl86}
D. H. Dunlap and V. M. Kenkre,
{\em Dynamic localization of a charged particle moving under the influence of an electric field},
Phys. Rev. B {\bf 34}, 3625 (1986).

\bibitem{Madi98}
K. W. Madison, M. C. Fisher, R. B. Diener, Qian Niu, M. G. Raizen, 
{\em Dynamical Bloch band suppression in an optical lattice}, 
Phys. Rev. Lett. {\bf 81}, 5093 (1998).

\bibitem{Lign07}
H. Lignier, C. Sias, D. Ciampini, Y. Singh, A. Zenesini, O. Morsch, and E. Arimondo,
{\em Dynamical control of matter-wave tunneling in periodic potentials},
Phys. Rev. Lett. {\bf 99}, 220403 (2007).

\bibitem{Ecka09}
A. Eckardt, M. Holthaus, H. Lignier, A. Zenesini, D. Ciampini, O. Morsch, and E. Arimondo
{\em Exploring dynamic localization with a Bose-Einstein condensate},
Phys. Rev. A {\bf 79}, 013611 (2009).

\bibitem{Long06}
S. Longhi, M. Marangoni, M. Lobino, R. Ramponi, P. Laporta, E. Cianci, and V. Foglietti,
{\em Observation of dynamic localization in periodically curved waveguide arrays},
Phys. Rev. Lett. {\bf 96}, 243901 (2006).

\bibitem{Sias08}
C. Sias, H. Lignier, Y. P. Singh, A. Zenesini, D. Ciampini, O. Morsch, and E. Arimondo,
{\em  Observation of photon-assisted tunneling in optical lattices},
Phys. Rev. Lett. {\bf 100}, 040404 (2008).

\bibitem{Ivan08}
V. V. Ivanov, A. Alberti, M. Schioppo, G. Ferrari, M. Artoni, M. L. Chiofalo, and G. M. Tino,
{\em Coherent delocalization of atomic wave packets in driven lattice potentials},
Phys. Rev. Lett. {\bf 100}, 043602 (2008).

\bibitem{Haller}
E. Haller, H.-C. N\"agerl, private communication.

\bibitem{Zhao91}
 X.-G. Zhao, 
{\em Dynamic localization conditions of a charged particle in a dc-ac electric field}, 
Phys. Lett. A {\bf 155}, 299 (1991).

\bibitem{remark0}
We stress one more time that in 2D tilted lattices the particle can be delocalized only in the direction orthogonal to the vector ${\bf F}$. In the direction parallel to ${\bf F}$ it is always localized with the Stark localization length $\sim t/aF$.

\bibitem{Tarr12}
L. Tarruell, D. Greif, T. Uehlinger, G. Jotzu, and T. Esslinger,
{\em Creating, moving and merging Dirac points with a Fermi gas in a tunable honeycomb lattice},
Nature {\bf 483}, 302 (2012).

\bibitem{Uehi13}
T. Uehlinger, D. Greif, G. Jotzu, L. Tarruell, T. Esslinger, Lei Wang, and M. Troyer,
{\em Double transfer through Dirac points in a tunable honeycomb optical lattice},
Eur. Phys. J. Special Topics {\bf 217}, 121 (2013).

\bibitem{92}
A.R.Kolovsky and E.N.Bulgakov,
{\em Wannier-Stark states and Bloch oscillations in the honeycomb lattice},
Phys. Rev. A {\bf 87},033602 (2013).

\bibitem{Naka93}
T.~Nakanishi, T.~Ohtsuki, and M.~Saitoh, 
{\em Stark ladders in two-dimensional tight-binding lattice}, 
J. of Phys. Soc. Japan {\bf 62}, 2773 (1993).

\bibitem{book}
Yu. A. Mitropolskii,
{\em Averaging methods in nonlinear dynamics},
Naukova Dumka, Kiev, 1971 (in russian).

\bibitem{remark1}
Eq.~(\ref{6}) as well as Eqs.~(\ref{10}-\ref{11}) below exclude the case where the vector ${\bf F}$ is aligned with vertical or horizontal bonds in Fig.~1. This case requires a separate consideration \cite{Sup}.

\bibitem{Sup}
See Appendix.   

\bibitem{Adel11}
M.~Aidelsburger, M.~Atala, S.~Nascimb\'ene, S.~Trotzky, Y.-A.~Chen, and I.~Bloch, 
{\em Experimental realization of strong effective magnetic fields in an optical lattice},
Phys. Rev. Lett. {\bf 107}, 255301 (2011).

\end{thebibliography}
\end{document}